\documentclass{article}

\begin{document}

\qquad SUPERSYMMETRIC YANG MILLS FIELDS AND BLACK HOLES:

\qquad \qquad \qquad IN TEN DIMENSIONAL UNIFIED FIELD THEORY

\qquad \qquad \qquad  AJAY PATWARDHAN

Physics department, St Xavier's college, Mumbai, India

Visitor, Institute of Mathematical Sciences, Chennai, India

\qquad \qquad \qquad \qquad ABSTRACT

The ten dimensional unified field theory has a 4 dimensional Riemannian space
time and six dimensional Calabi Yau space  structure. The Supersymmetric Yang
Mills fields and black holes are solutions in these theories. The
formation of 
primordial black holes in early universe, the collapse to singularity
of stellar black holes, the Hawking evaporation of black holes, the possible
creation and evaporation of microscopic black holes in LHC are topics
of observational and theoretical interest.

 The observation of gamma ray bursts and creation of spectrum of particles and radiation of dark
and normal matter occur due to primordial and microscopic  black holes.
The approach to singularity in black hole interior
solutions, require the Bogoliubov transforms of SUSY YM fields in
black hole geometries; both during formation and in evaporation. The
Hawking effect of radiating black holes is applicable for all the
fields. Invariants can be defined to give the conditions for these processes.

\qquad \qquad \qquad \qquad  1. INTRODUCTION :

Yang Mills fields in black hole geometries is a subject that has a revived interest. Primordial black holes created in the early universe can have quantum evaporation with Yang Mills fields, charges and particle creation. Observation of isotropic distribution of gamma ray bursts and creation of normal and supersymmetric particles (possibly normal and dark matter), have renewed need of a supersymmetric Yang Mills theory in black hole physics. 

Black holes formed by stellar collapse following a supernova, and accretional processes of black holes and their mergers, also raise possibilities for remnant Yang Mills contributions. The nature of the singularity in black holes, needs a detailed study with SUSY YANG MILLS theory.

The 10 dimensional unified field theory developed in my earlier papers, has implications for cosmology as well as for black hole physics. The presence of topological invariants in SUSYYM and the covariant description of black hole formation and evaporation, can give the necessary analysis for additional charges on black holes and conditions on the singularity. The black hole emerges as a true thermodynamic engine for matter creation and extinction, a geometric effect in the unified field.

Primordial black holes and stellar collapse black holes are both considered. At the end stage of collapse an equation of state of the matter is required, the geometry of the interior solution is like a cosmological model, with time and radial coordinates exchanged.
The relativistic and high temperature limit of fermionic and bosonic partition functions and equations of state are of a quantum ideal gas
with a $ h^3 $ minimum volume in phase space per particle. Asymptotic freedom of quarks at very short separations creates a quantum gas of very weakly interacting and essentially free quarks. 

Even though degeneracy pressure and exclusion principle for quarks and
leptons is unable to halt gravitational collapse to a stable star
state, the quantum ideal effectively spin less gas has a statistical
ensemble minimum volume limit:$ N h^{3} $.  with $ N = 10^{60} $ and
the number of independent configurations will be $ h^{3N} $. The picture is of N particles clustered around the origin occupying the minimum possible volume sphere, which is the size of a nucleus, radius of sphere =$ 10^{-14}$ meters. There is no singularity at the origin in the sense of infinite density or curvature, although both are a very large number. 

Conservation of the Yang Mills fluxes and charges of the interacting particles and fields in a covariant description of the collapse, creates a finite number of Yang Mills charge. While this YM charge should be included in the interior solution for the geometry inside the horizon for consistency of the collapse calculation, the presence of YM 'hair' outside the horizon as a charge occurring in the exterior solution is predicted by some papers and refuted by others. 

Thus the Kerr Newmann solution modified to include the YM charges requires a quantum effect to be included in a classical geometry. The YM fields of the unified field theory, other than electromagnetic, are all quantum fields with quantum charges. Hence this may not work for stellar collapsed classical black holes with horizon radii in kilometers. 

The primordial black holes have other mechanisms of formation. In this
case the mass spectrum and hence the horizon radii scales are spread
over a large range. Quantum effects dominate in the formation and
evaporation for primordial black holes of mass $ < 10^{ 15} $ kg. The YM charges and their effects on this process should appear in the interior and exterior geometry. SUSYYM theory allows for exchange of fermionic and bosonic degrees of freedom. This is done by Bogoliubov transforms that depend on the spacetime geometry. 

In the end stage of collapse the distributions for fermionic and bosonic gas coincide and the SUSY YM with Bogoliubov transforms can be shown to make a consistent theory of the matter and of the geometry in the interior solution of the black hole. In the case of evaporation by Hawking radiation of primordial black holes there is no static solution of the spacetime metric, and the SUSY YM charges are included in the geometry. The decay of mass, angular momentum, EM charge, YM charge, yields the particle spectrum generated. 

What went in and what came out of the primordial black hole have only global covariant conservation laws governing them. The rest is subject to transformations. If the particles created in this quantum evaporation contribute to a significant amount of the content of the observable universe, then the exact theory for this process has cosmological significance. However if such events are only locally of interest then they will be a phenomenon to explain in astronomy. Another question is of normal and dark matter in collapse and in evaporation. The SUSY YM unified field theory in a dynamic geometry, including Bogoliubov transforms, allows transformation of all these forms of matter.

The usual theories of black hole formation with interior solution use variations of Vaidya, Raychowdhury and Friedman equation; depending on properties of the fluid matter used. In the analysis of the unified field theory it is possible to define partition functions and phase transitions that yield Calabi Yau spaces with the most weightage dependent on Euler number; hence simple geometries are formed. YM fluxes and charges are covariantly conserved and finally give the description of the neighbourhood of the origin. 

String theory gives a "ball of twine" description, non commutative geometry has a "fuzzy sphere", quantum gravity has " polymer network" and "foam" description of space time and matter around  the origin. While these regimes may occur in the interior of a black hole, it is the prediction and verification of observable facts that will decide. Hawking and Penrose theorems for singularities, while being quite general in their geometric form are classical and do not include quantum effects. They are actually not equipped to state the nature of the neighbourhood of the origin. 

The unified field theory includes gravity and hence the gravitational
radiation modes and the static geometries too. The picture for
applying this theory to the evolution of the universe was that the ten
dimensional theory gives a four dimensional Riemannian space time
described by Einstein's theory and a compactified Calabi Yau space for
the $ U(1) \otimes SU(2)\otimes SU(3) $ Yang Mills theory. This occurs as the energy scales go down from Planck scale.

 In the end stage of collapse, conversely, the energy scales go up towards Planck scale and the unified theory combines all interactions with gravity. The neighbourhood of the origin is therefore Calabi Yau spaces of  compact size and six dimensions wrapped around the four dimensional spacetime origin. The quantum statistical fluctuations and averages of the quantities in this theory give the picture of a very active region around the origin; not simply a "point" singularity.

 If the collapse of a stellar mass black hole had taken electrons,
 nuclei, neutrons down below the horizon and driven them into the
 origin; around $ 10^{60} $ particles would have fallen into the origin and
 clustered around it. Quarks and leptons as a asymptotically free
 quantum gas would have occupied the least allowed volume around the
 origin. In the unified field theory this picture is possible in the
 sense of the SUSY YM on Calabi Yau 6 dimensional spaces which wrap
 around the origin of the 4 dimensional space time, with $ 10^{60} $
 particles sticking out like pins on a pin cushion or thumbtacks on a
 crumpled paper ball.

The scalar fields in inflation generate the SUSY YM particle spectrum in the
expanding geometry of spacetime. They also have local fluctuations
collapsing to form primordial black holes. the quantum evaporation of
the PBH gives rise to a thermal like spectrum corresponding to black
hole temerature and all SUSY YM  particles are 
emitted with rest energy corresponding to temperature of surrounding
region. The scalar field particles could be the dark matter.

\qquad \qquad I BOGOLIUBOV TRANSFORMS FOR SUSY YM FIELDS :
\qquad \qquad IN  BLACK HOLE'S  INTERIOR AND EXTERIOR GEOMETRIES

For simplicity the interior geometry is a Schwarzschild, Kerr Newmann,  Friedmann Robertson
Walker or Bianchi metric
with the radial and time  coordinates swapped.
The exterior is a Schwarzschild or Kerr Newmann metric near the
horizon and the cosmological model of FRW or Bianchi metric in the distance. For all
these geometries there are standard Bogoliubov transform expressions
and the $ \beta \beta^{\dagger}$ or number operator has a known form.

The transformation defined as follows allows analytic continuation of
the solution across the horizon:

For the supersymmetric fields the fermionic and bosonic operators are
in a representation with transforms as follows:

The creation and anhilation operators $f^{\dagger}$ and $f$ transform as

$ \bar {f} = \alpha f +\beta   f^{\dagger}$ ; 

  and similarly the  adjoint case, and for the bosonic $b$ operators . 

This transform is anti  commutator  and commutator 
algebra preserving for the conditions

$\mid \alpha \mid^{2}  \pm \mid \beta \mid ^{2} = 1 $

for Fermi$ (+)$ and Bose$ (-)$  sign cases

The number operator is $b^{\dagger}b$ for bosonic and  $f^{\dagger} f $ for fermionic case and

 its expectation value is 

$<0\mid N \mid 0> = = \mid \beta \mid^{2}$

 More generally the $\alpha $ and $ \beta$ can be matrices for multimode systems. The transforms with the $q$ deformed commutation were also given in [Ref].

The traceclass and Hilbert Schmidtt  condition on the $\beta \beta^{\dagger}$ is required. 

A more general Bogoliubov transform can be defined that acts on the supersymmetric matrix operators above and can give distinct as well as mixed sectors as special cases.

\begin{displaymath}
\left(\begin{array}{c|c}
\phi & \psi \\
\bar \psi & \bar\phi
\end{array}\right)
\end{displaymath}

and its inverse can be defined to make the Bogoliubov transforms invertible.

$ \psi  \psi^{\star} $ traceclass, Hilbert Schmidtt

 and $\phi  \phi^{\star} $ has a Fredholm determinant 

give a proper canonical transform iff 

$ \phi \phi^{\star} = I + \psi  \psi^{\star} $

For matrix operators $\phi$ and $\psi$ this property becomes

$\phi\phi^{\dagger} = I + \psi \psi^{\dagger}$ 

and $\phi \psi^{tr} = \psi \phi^{tr} $

With these the system of many bosonic and fermionic variables can be treated.

The Calabi Yau spaces for the SUSY YM fields are taken as multi
polynomial forms in  compactified 6 dimensions  where all the
relevant symmetries and particle states are included.

The invariants of the SUSY YM fields are written as: Integrals of $ *F $, $ F $, $ *F \wedge *F
 $, $ F \wedge F $.

 The Yang Mills charges contribute to the metric and all
 the gauge and general covariant quantities give conserved geometric
 and topological indexes.

In the Robertson Walker and Bianchi I geometries the mode functions $\chi_{k} (\eta)$ transform and give the Bogoliubov transform coefficients

The $\beta $ factors  in the Bogoliubov transforms give the condition on number operators to have expectation values given by

$\mid\beta\mid^2 = \frac {Sinh^{2}(\frac{(\pi \omega_{-})}{\rho})}{Sinh(\frac{\pi \omega_{i}}{ \rho}) Sinh(\frac{\pi \omega_{f}}{ \rho})} $

where  $ \omega_{\pm} = 0.5(\omega_{i} \pm \omega_{f} ) $ ;

and $\omega_{i} $ is  frequency at initial time and $\omega_{f} $ is
frequency at final time.

For the internal geometry the timelike Killing vector is replaced by a
spacelike one, the energy operator by the momentum one. The radial
momentum operator with eigen modes given by conditions at horizon
radius and at origin are used to find the Bogoliubov coeffecients. The
analysis carried out for external geometry, uses the timelike vector
and energy eigen modes. The positive and negative frequency solutions
are now replaced by outgoing and incoming modes along the radial
direction. And the transforms between horizon and origin are taken.

The vaccuum states are constructed at horizon and at origin ( for
formation and
collapse) , and at
the horizon and at infinity ( for the evaporation). The
condition of $t_{in} $  to $ t_{fin} $ during evaporation is  replaced by $r_{h} $ and $
r_{0} $
 during collapse and formation. The role of $ t $ and $ r $ are swapped.

The scale factor$ a(t) $ of the geometry $ dS^{2} = dt^{2} - a^{2} (t) dx^{2}$

 with $\eta = \int\frac{1}{a(t)} dt$

and the $C(\eta ) = a^{2} (\eta)$ with

 $ \int a(\eta) d \eta = \int^{t} dt$ as the scale factor

 and $C(\eta) = a^{2}(t).$

 Then $ c(\eta) = A + B  tanh (\rho \eta) $ becomes $ A \pm B $ as $ \eta $ becomes $\pm $ infinity. 

Introducing three parameters $ A, B, \rho $ we get the mode functions that satisfy the equations

 $\frac{d^{2} \chi_k} {d \eta^{2} } + (k^{2} +  C(\eta) m^{2} ) \chi_k (\eta) = 0$

giving  $\omega_{i} = ( k^{2} +m^{2} (A -B))^{1/2} $ 

and $\omega_{f} = ( K^{2} + m ^{2} ( A +B ) )^{1/2}$

It is these equations that can be compared with the supersymmetric Hamiltonian , with the $W^{2}(\eta)$ and $\frac{dW(\eta)}{d\eta} $ terms. In that notation the collection of all operators $b , f, Q, H $ and their adjoints were constructed and now these have a specific realisation in terms of the field mode function equation in the curved space time geometry.

 These mode functions allow construction of the coefficients for the Bogoliubov transforms on the operators. 

For example, $b(t_{f}) = \alpha_{(t_{f},t_{i})}  b(t_{i}) +  \beta_{(t_{f},t_{i})}^{\star} b^{\dagger}(t_{i})$

Then the $\beta \beta^{\star} (t_{f},t_{i} )$ gives the non zero expectation values and particle creation.

Following [REF ] this gives the required $\mid \beta \mid^{2} $ as above. 

The particle creation rate can be found. Depending on the Fermi or Bose fields used these expressions are found for normal and supersymmetric partners.

In the cosmological models with a $10^{-5}$ anisotropy as seen from WMAP data, the Bianchi I model with slight anisotrpic case gives the same expression as above with the $m^{2} $ term modified by

 $ \alpha^{2} cosec^{2} (\eta) $ where $\alpha$ is the anisotropy parameter.

 A similar expression will be used for slightly anisotropic collapse in interior solution of a black hole geometry.

Similar to the number operator  expectation values , we can also find the  vaccuum expectation values of components of the stress energy tensor

$\frac{ <0(f)\mid T_{\mu \nu} \mid  0(i)> }{<0(f) \mid 0(i)>}$ 

between any initial and final time vaccuum states in the geometry.

\qquad\qquad II PATH INTEGRAL PARTITION FUNCTION FOR UNIFIED FIELD THEORY:

The partition function $ Z(M) =\int d[\phi]Exp[-S(\phi)]   $

 is modified to  $\Sigma dim(M)^{\chi(M)}\int d[\phi_i,A_i]exp[-S(\phi_i,A^a_i)] $

 and the Holonomy is given as  $ TrPExp(-ie\oint_ {M(c_i)} dx A^at_a ) ) = W[M_n,A^a_i,C_i]$

 For the compactified subspaces the partition function becomes 

 $ Z(M_i) = \Sigma_i dim (M_i)^{\chi(M_i)} \int_{M_i} d[\phi,A] Exp( -S[\phi,A])$

The compactifications could be expressed in the Kaluza Klein way as  a line element, to illustrate the concept,

$  ds^2 = f(r,t)dt^2 -g(r,t)dr^2 -\Sigma_{\mu,\nu}\Sigma^{dimG(i)} (A^a_\mu  A^b_\nu t_a t_b)_{ij} \zeta^i_\mu \zeta^i_\nu  $;

 $\mu , \nu = 3 $to $ 10 $.

The partition function is expressed as

 $ Z = \Sigma_{M(n)=1}^{10} dim (M(n)^{\chi(M(n))} \int_{M(n)}d[\phi_i,A^a_j] Exp(-S[\phi_i,A^a_j]) $

and the expectation values of the Holonomies are taken as

 $ <W[(C)]> =i/Z \int_{G} D[A] W[C,A] exp(-S_{YM}(A))$ on loops $C$

the general result is

$<W(C_!)--W(C_n)>=1/Z \int D[A](W(C_1,A) --W(C_n,A)) Exp(-S_{YM}(A)) $

for the n loops Holonomy  and the  $-1$ has been inserted in the exponent as  replacement of the $2\pi i/h$ as coeffecient of the Action 

$ \int(\phi_j^+ \phi_k) exp(-\phi^+ F \phi)d\phi^+d\phi  = (\int exp(-\phi^+ F \phi)d\phi^+d\phi )) 1/F_{jk}$

 gives for quadratic forms, gaussian like integrals which could be used for model calculations

\qquad \qquad III FORMATION AND EVAPORATION OF BLACK HOLES ( STELLAR, PRIMORDIAL AND
MICROSCOPIC) IN SUSY YM THEORY:

The fields are the SUSY YM and scalar fields. During collapse the
accelerated frame Bogoliubov transforms in the interior geometry, from
vaccuum at horizon to vaccuum at origin require a quantum treatment of
the in fall of matter. The number operator expectation values are
different when taken at the horizon,  which is the normal state of the
star collapsing and the
approach to singularity, which leaves a black hole in spacetime. Hence
the number is augumented by the factor $ N_{star} \beta\beta^{\dagger}
$

For the bosonic system the transform is 

\begin{displaymath}
\left(\begin{array}{c}b^{'}(k)\\b^{'}(-k)^{\dag} \end{array}\right) = 
\left(\begin{array}{cc}cosh(g(k,a)) & sinh(g(k,a))\\ sinh(g(k,a))& cosh(g(k,a))\end{array}\right)\left(\begin{array}{c}b(k)\\b(-k)^{\dag}\end{array}\right)\end{displaymath}

Similar expressions for the fermionic system are

\begin{displaymath}
\left(\begin{array}{c}c^{'}(k)\\c^{'}(-k)^{\dag} \end{array}\right) = 
\left(\begin{array}{cc}cos(g(k,a)) & sin(g(k,a))\\- sin(g(k,a))& cos(g(k,a))\end{array}\right)\left(\begin{array}{c}c(k)\\c(-k)^{\dag}\end{array}\right)\end{displaymath}

The Number operator has vaccuum expectation values

 $ sinh^{2}(g(k,a)) $ and $ sin^{2}(f(k,a)) $

The entropy of the black hole is $ S = K Ln \Omega $ and the number of
degrees of freedom and their independent  configurations scale as the
number of independent solid angles, into which the particles arrange
themselves. These project onto the horizon surface as area
patches. Hence entropy varies as number of these patches, covering the
horizon area, hence the area theorem: entropy is proportional to the
area of horizon.

For the formation of  primordial black holes, the inflationary scalar field has
quantum fluctuations in $ < \phi^{4} > $ which can create horizon
condition for PBH. The energy density of the scalar field is a
partition function average, and gives the parameters of the PBH. The
other fields of the unified field coupled to the scalar field have a
particle creation rate depending on the decay of the scalar field
during the inflationary expansion of the cosmological model.[Ref]

For the quantum evaporation of the PBH, the unified field has '
quantum hair' and Hawking radiation of all allowed particle
states. The PBH radiates mainly when its temperature (inverse mass) is
comparable or greater than the background radiation / matter in the
universe; and then most probable emission is of particles whose rest
energies are of the order of  back ground thermal energies. The range
of masses of PBH from $10^{-18} $ kg to $ 10^{12} $ kg decayed
earliest and the gamma bursts of these objects were not observed.

 The observable horizon (light just reaching us now); has PBH of mass  $ >
10^{12} $ kg  are decaying and these gamma ray  bursts accompanied by particle
creation are observed.  PBH of mass $ 10^{12} $ kg to $ 10^{19} $ kg
are also formed but they have a longer lifetime and will decay later,
slower and with lower black body temperatures. Their radiation will be
observable far in the future.

The SUSY YM theory coupled to the scalar inflation field accounts for
the dark matter in uniformly distributed background. This dark matter is not
coupled to electromagnetic radiation. The PBH evaporation from local events
distributed in the early universe accounts for the normal and
supersymmetric partners creation. These two mechanisms result in a $ 5
: 1 $  ratio of dark and normal matter as well as the $ 10^{-5} $
 inhomogenity in the universe.

 The PBH acts like  a thermodynamic
 engine converting the scalar field's energy fluctuation during formation into the
 energy density of normal and supersymmetric matter and
 electromagnetic radiation during evaporation. The unified field
 theory of SUSY YM and gravity,  which is four dimensional Riemannian space time and six dimensional
 Calabi Yau spaces contains black hole solutions with classical and
 quantum hair.

\qquad \qquad \qquad\qquad CONCLUSION

The reference list indicates the diverse work going on in the
subject. The particular and distinct results developed in papers have been related to
the unified gauge field theory in ten dimensions of my previous work. 
It is clear that within a few years due to precise observations in
cosmology and from the large hadron collider at CERN, it will be
possible to get the necessary facts to put the pieces of the puzzle
together into a consistent framework. This series of arxiv eprints by
me is an attempt in that direction.

The internal and external geometry of black holes, their horizons and
singularity, admit a SUSY YM theory, with Bogoliubov transforms. The
formation and evaporation of primordial black holes in the early
universe and microscopic black holes in the laboratory will require
some new physics and have significance.

The unified field theory has the possibility of producing this new physics.

\qquad \qquad \qquad ACKNOWLEDGEMENTS 

I thank Prof H S Sharatchandra and Prof Balasubramaniam, Director of
Institute of Mathematical Sciences , Chennai for their support for my
visit. Discussions with Prof Sharatchandra, Prof Date and others are acknowledged.

\qquad \qquad \qquad REFERENCES

1.Phys Rev D 75,024022 (07)

2.Phys Rev D 60, 063516 (07), Anne Green

3.Phys Rev D 75, 044006 (07)

4.gr-qc/0510043 Farley, d'Eath

5. astro-ph/9903484

6. QFT in curved spaces, Birell, Davies 

7.C.Kieffer Quantum Gravity

8.quant-ph/0305150 Ajay Patwardhan

9.hep-ph/0405215 J Feng

10.hep-th/0406049 Ajay Patwardhan

11.gr-qc/0510027 Farley, d'Eath

12.hep-th/0606080 Ajay Patwardhan

13.hep-th/9904006 P Mazumdar

14.Phys Rev D 65, 027301 Anne Green

15. Phys Rev D 68,044016

16.Phys Rev D ,62, 044046

17.Phys Rev D 65, 061502

18.Phys Rev D, 59, 124013

19. Phys Rev D 40 , 1748

20.PRL, 69,13,1852

21. PRL, 78,18,3430

22. Phys Rev D 56, 6,3459

23. PRL, 86,17,3704

24.Phys Rev D 52,10,5659

25. Phys Rev D 71,104009

\end{document}